\documentclass[prc,twocolumn,showpacs,amsmath,amssymb]{revtex4}

\usepackage{graphicx}
\usepackage{dcolumn}
\usepackage{bm}

\begin{document}

\title{Shell corrections for finite depth potentials with bound states only}

\author{A. Diaz-Torres}
\email{diaz@th.physik.uni-frankfurt.de}
\affiliation{Institut f\"ur Theoretische Physik der
Johann Wolfgang Goethe--Universit\"at Frankfurt, Robert Mayer 10,
D--60054 Frankfurt am Main, Germany and \\
Institut f\"ur Theoretische Physik der
Justus--Liebig--Universit\"at Giessen, Heinrich--Buff--Ring 16,
D--35392 Giessen, Germany}

\date{\today}

\begin{abstract}
A new method of calculating unique values of ground-state shell
corrections for finite depth potentials
is shown, which makes use of bound states only. It is based on (i) a general
formulation of extracting the smooth part from any fluctuating quantity proposed by Strutinsky
and Ivanjuk, (ii) a generalized Strutinsky smoothing condition suggested recently by Vertse et al.,
and (iii) the technique of the Lanczos $\sigma$ factors. Numerical results for some
spherical heavy nuclei ($^{132,154}$Sn, $^{180,208}$Pb and $^{298}$114)
are presented and compared to those obtained with the Green's function oscillator expansion method.
\end{abstract}



\maketitle


The binding energies of nuclei as a
function of nucleon numbers and shape parameters show an irregular behaviour superimposed on a smooth trend.
This irregularity (shell corrections) is mainly due to the gross-shell non-uniformity of the single-particle
(sp) levels near the Fermi energy. The smooth trend can be reproduced well by the liquid-drop or droplet
mass formula \cite{Myers}, while the shell corrections can be calculated with the Strutinsky method
\cite{Struky1,Brack}.

The Strutinsky method converges very well (there is an interval of values \cite{Struky2} of both
the energy smoothing parameter and the order of the curvature-correction polynomial where the shell
correction practically remains the same) when it is applied to infinite depth potentials like the harmonic
oscillator potential or a Nilsson model potential. Difficulties appeared \cite{Struky2}, however,
in calculations using finite depth potentials where
the number of bound levels above the Fermi energy was not enough for the application of the traditional
energy-averaging ($E$-averaging) Strutinsky procedure \cite{Struky1,Brack}.

In several papers, e.g., Refs.
\cite{Bolsterli,Ross,Brack2}, ideas were discussed to improve the convergence of the $E$-averaging procedure for finite potentials by including the continuum
effect on the level density. Other methods \cite{Bhaduri1,Jennings1,Sobiczewski1} have also been applied to obtain shell
corrections of finite potentials. Very recently in Ref. \cite{Vertse1}, the authors have drawn the conclusion that the proper treatment of the continuum effect on level density
does not guarantee that the standard $E$-averaging procedure converges.
Consequently, an alternative prescription for defining shell corrections of finite potentials, based on the assumption of a
local linear energy dependence of the smoothed level density (generalized Strutinsky smoothing condition),
was suggested. Moreover, a practical method of calculating the
smooth sp level density in combination with the proposed definition of shell
correction was used in Ref. \cite{Vertse2} to obtain unambiguous shell corrections of the finite deformed Woods-Saxon
potential for drip-line nuclei. This new procedure was called Green's Function Oscillator Expansion
(GFOE) method.

Apart from the methods mentioned above, there are prescriptions based on averaging in the nucleon
number space ($N$-averaging) \cite{Struky3,Ivanjuk1,Ivanjuk2,Ivanjuk3,Tondeur}.
The $N$-averaging method does not need any unbound levels. The difference between the $N$-averaging
and the $E$-averaging procedures is due to the symmetry of the sp hamiltonian, i.e., the degeneracy of the sp levels \cite{Ivanjuk2}.
The $N$- and $E$-averaging are particular cases of a general
formulation of extracting the smooth part from any fluctuating quantity, suggested by Strutinsky and
Ivanjuk in Ref \cite{Struky3}. Unfortunately, the versions of the
$N$-averaging prescription discussed so far in the literature
(e.g., Refs. \cite{Struky3,Ivanjuk1,Ivanjuk3}) do not seem to yield stable results.

In this paper a new method of calculating the ground-state shell
corrections of finite depth potentials is proposed,
which makes use of bound states only. The method is based on (i) the general formulation shown in
Ref. \cite{Struky3}, (ii) the generalized Strutinsky smoothing condition suggested recently in
Refs. \cite{Vertse1,Vertse2}, and (iii) the technique of the Lanczos $\sigma$ factors \cite{Lanczos}.



$\textit{Method:}$ (i) In the general formulation presented in Ref. \cite{Struky3}
(when it is referred to the nucleon number space, $N$- averaging) the smooth part
$\bar{E}_{s}(N)$ of the total sp energy sum $E_{s}(N)$
($N$ is the nucleon number of the system ) was determined as the least-square-deviation (LSD)
fit to $E_{s}(N)$ by an M-th
degree polynomial of $N^{1/3}$ \cite{Struky3,Ivanjuk1,Ivanjuk2,Ivanjuk3}.
The smoothed total sp energy $\bar{E}_{s}(N)$ is expressed as

\begin{equation}
\bar{E}_{s}(N)=\sum_{\lambda=N_1}^{N_2} \zeta_M (N,\lambda)E_{s}(\lambda), \label{eq_1}
\end{equation}
where $N_1$ and $N_2$ are the lower and the upper limits of the averaging interval
$N_1 \leq \lambda \leq N_2$. From now on we will
fix $N_1$ and $N_2$ to the minimum and the maximum nucleon number that the potential-well can contain, i.e.,
$N_{1}=1$ and $N_{2}=N_{max}$ which is the sum of the degeneracy of all individual bound states.
It is important to point out that the degeneracy of a sp level will not be taken into account in (\ref{eq_1})
with a degeneracy
factor, because the nucleons numbered by $\lambda$, belonging to a degenerated sp level,
appear with different weights $w(\lambda)$ in the smoothing function $\zeta_M$ which reads as

\begin{equation}
\zeta_M (N,\lambda)=K_{M}(N,\lambda)w(\lambda). \label{eq_2}
\end{equation}

The curvature-correction $K_M(N,\lambda)$ is the polynomial of the M-th degree, which is composed
of polynomials $p_{k}(\lambda)$ orthonormal in the discrete interval
$1 \leq \lambda \leq N_{max}$ with the weight $w(\lambda)$ \cite{orthogonal_poly} and
reads as

\begin{equation}
K_{M}(N,\lambda)=\sum_{k=0}^{M}p_{k}(N)p_{k}(\lambda), \label{eq_3}
\end{equation}
where $\sum_{\lambda=1}^{N_{max}}p_{k}(\lambda)p_{k^{'}}(\lambda)w(\lambda)=\delta_{kk^{'}}$.
The smoothed total sp energy $\bar{E}_{s}(N)$ in expression (\ref{eq_1}) can be written in terms
of the sp energies $e_{\lambda}$ and the smoothed sp occupation numbers $\bar{n}_{\lambda}$ as

\begin{equation}
\bar{E}_{s}(N)=\sum_{\lambda=1}^{N_{max}} e_{\lambda}\bar{n}_{\lambda}, \label{eq_4}
\end{equation}
where $\bar{n}_{\lambda}=\sum_{\nu=\lambda}^{N_{max}} \zeta_M (N,\nu)$ and
$\sum_{\lambda=1}^{N_{max}}\bar{n}_{\lambda} = N$.
The conservation of the number of particles is guaranteed by the definition of the smoothing function
$\zeta_M$ (\ref{eq_2})-(\ref{eq_3}).
The argument of the smoothing function $\zeta_M$ is understood as $x_{\lambda}$ rather than $\lambda$, where the
variable $x_{\lambda}$ is a linear function of $\lambda^{1/3}$, i.e.,
$x_{\lambda}=(\lambda^{1/3} - N_{0}^{1/3})/\Delta$.
The parameters $N_0$ and $\Delta$ determine the maximum and the width of the weight
function $w(x_{\lambda})$ which will be a Gaussian
$w(x_{\lambda})=\exp(-x_{\lambda}^2)/\sqrt{\pi}$ centered at the Fermi level, i.e.,
the parameter $N_0$ is fixed at the nucleon number $N$ for which the shell correction in the
$N$-averaging procedure

\begin{equation}
\delta U(N)=E_{s}(N) - \bar{E}_{s}(N) \label{eq_7}
\end{equation}
is calculated. Therefore, the remaining free parameters of the equations above are $\Delta$
and the
$M$ degree of the smoothing function $\zeta_M$. An unambiguous (converged) value of the shell
correction
$\delta U(N)$ means that it should not depend (or very weakly at most) on
$\Delta$ and $M$, which will be discussed below.

(ii) In our calculations we have not found any plateau of $\delta U(N)$ regarding the parameter $\Delta$,
for a fixed $M$ degree of the smoothing function $\zeta_M$. The value of $\Delta$
is searched around $\Delta = 1$, which approximately corresponds to the energy smoothing parameter
($\gamma = 1.2 \hbar\omega_0$) of the traditional $E$-averaging procedure,
$\hbar\omega_0 = 41 A^{-1/3}$ MeV being the mean distance between the gross-shells. Consequently,
we follow the generalized Strutinsky smoothing condition suggested in Refs. \cite{Vertse1,Vertse2}
to obtain an optimal value of the parameter $\Delta$ ($\Delta_{op}$) for a given $M$.
This condition was supported by extensive calculations \cite{Vertse1,Vertse2} of the smoothed sp level
density using both the semiclassical Wigner-Kirkwood method and the exact expression of the level
density which incorporates the continuum effects properly. From these studies, it was observed that
the smoothed level density shows a linear energy dependence in the intermediate energy region of bound
states. In Refs. \cite{Vertse1,Vertse2}, the following generalized smoothing condition was
suggested: In an energy
interval $[\epsilon_l,\epsilon_u]$ which is wider than the mean distance $\hbar\omega_0$ between
the gross-shells (e.g., $\epsilon_u - \epsilon_l = 2\hbar\omega_0$), the deviation of the smoothed
sp level density $\bar{g}(e,\Delta,M)$ from linearity should be minimal. Hence, $\Delta_{op}$ for a
given $M$ is calculated by minimizing

\begin{equation}
\chi^2 (\Delta,M)=\int_{\epsilon_l}^{\epsilon_u}[\bar{g}(e,\Delta,M) - a -be]^2 de, \label{eq_smooth1}
\end{equation}
where the parameters $a$ and $b$ are uniquely determined for each value of $\Delta$ and $M$ by the method
of least squares. The smoothed sp level density $\bar{g}(e,\Delta,M)$ is obtained as follows:
The smoothed total sp energy $\bar{E}_{s}(\lambda)$ ($\lambda$ refers to the number of particles
with which the sp levels are filled, i.e. $1 \leq \lambda \leq N_{max}$)
can be written in terms of $\bar{g}(e)$ as

\begin{equation}
\bar{E}_{s}(\lambda)=\int_{-\infty}^{\bar{\mu}_{\lambda}} e\bar{g}(e)de, \label{eq_10a}
\end{equation}
with $\bar{\mu}_{\lambda}$ determined by the condition of conservation of the number of particles

\begin{equation}
\lambda=\int_{-\infty}^{\bar{\mu}_{\lambda}} \bar{g}(e)de. \label{eq_10b}
\end{equation}

From these two equations, $\bar{g}(\bar{\mu}_{\lambda})$ and $\bar{\mu}_{\lambda}$ can be
calculated as
\begin{equation}
\bar{g}(\bar{\mu}_{\lambda})=(\frac{d^{2}\bar{E}_{s}(\lambda)}{d\lambda ^2})^{-1}, \label{eq_10c}
\end{equation}
\begin{equation}
\bar{\mu}_{\lambda} = \frac{d\bar{E}_{s}(\lambda)}{d\lambda}, \label{eq_10d}
\end{equation}
where $\bar{E}_{s}(\lambda)$ are obtained with the moving average (\ref{eq_4}).
The expressions (\ref{eq_10c})-(\ref{eq_10d}) allow the construction of the function
$\bar{g}=\bar{g}(e,\Delta,M)$. The energy interval
$[\epsilon_l,\epsilon_u]$ is centered around the half of the energy of the lowest
sp level (i.e., $\sim 0.5e_1$). The search for $\Delta_{op}$ begins at a small $\Delta$ value below $\Delta = 1$, e.g., $\Delta = 0.5$
where the effect of the shell fluctuations on $\bar{g}(e,\Delta,M)$ is still present. $\Delta$ is
then gradually increased until the first minimum in $\chi^2 (\Delta,M)$ appears at $\Delta_{op}$. This
$\Delta_{op}$ represents the smallest value of $\Delta$ for a given $M$ that smooths out the shell
fluctuations effect on $\bar{g}(e,\Delta,M)$. The shell correction $\delta U(N,\Delta_{op},M)$
is the optimal one for a given $M$. This prescription
is repeated for higher values of $M$. If variations of $\delta U$ with $M$ are small
(e.g., $\leq 0.2$ MeV), then the mean value of those $\delta U$ represents a unique value of the
shell correction. The improvement of the convergence with respect to $M$ will be discussed
next.

(iii) Numerical instabilities happen concerning the convergence of the shell corrections with increasing
$M$ degree of the smoothing function $\zeta_M$. This is due to the effect of Gibbs
oscillations shown by Lanczos in the theory of applied Fourier-analysis in Ref. \cite{Lanczos}, which always
appear when a smooth function is expanded in terms of a complete orthonormal
set of functions and the expansion is truncated at a finite number $\textmd{N}$ of terms.
A technique to strongly
decrease the instabilities and accelerate the convergence of the shell corrections may use the
so-called Lanczos $\sigma$ factors \cite{Lanczos,Revai1,Revai2}.
The main idea of this method consists of damping the higher order
terms of the function expansion multiplying the expansion coefficients by attenuation factors
$\sigma_{k}^{\textmd{N}}$ ($0\leq \sigma_{k}^{\textmd{N}} \leq 1$).
The factor $\sigma_{k}^{\textmd{N}}$ is 1 only for the first value of $k$, e.g., $k=0$;
for increasing $k$ the $\sigma_{k}^{\textmd{N}}$ decreases monotonously and become almost zero for
the highest subscript $k=\textmd{N}$. The $k$-dependence of $\sigma_{k}^{\textmd{N}}$ is
arbitrary and can be chosen to fit the actual problem best.

The $\sigma$ factors
enter the shell correction method discussed above through expression (\ref{eq_3}) for the
curvature-correction polynomial $K_M$. For a given nucleon number $N$, expression (\ref{eq_3})
can be identified as a truncated expansion of the true curvature-correction function in terms of the
orthonormal polynomials $p_k (\lambda)$, where the expansion coefficients are $p_k (N)$. Therefore,
the modified coefficients $\sigma_{k}^{M} p_k (N)$ will replace the coefficients $p_k (N)$.
As in Refs. \cite{Revai1,Revai2}, the following $\sigma_{k}^{M}$ factors

\begin{equation}
\sigma_{k}^{M}=\frac{1-\exp \{-[\alpha (k-M-1)/(M+1)]^2 \}}
{1-\exp(-\alpha^2)}, \label{eq_16}
\end{equation}
are used, where the value $\alpha = 5$ was found in trials as in Refs. \cite{Revai1,Revai2} to
optimize the convergence rate, i.e., the formation of a plateau.

$\textit{Numerical results:}$ The bound states used in the present calculations are obtained with
the Potential Separable Expansion (PSE) method proposed by Revai in Ref. \cite{Revai1}.
The PSE's rigorous foundation was demostrated in Ref.
\cite{Revai2} and its efficiency was shown in several papers, e.g., Refs. \cite{Revai1,Revai3}.
For the sake of simplicity, the finite depth nuclear potential is chosen to be the following
spherical Woods-Saxon with a spin-orbit term
\begin{equation}
V(r)=-V_0f(r) + \frac{1}{2}\lambda (\frac{\hbar}{mc})^2 V_0 \frac{1}{r}\frac{df_{so}}{dr}
(\textbf{l}\cdot \textbf{s}), \label{eq_15}
\end{equation}
where $\hbar /mc = 0.21$ fm, $f$ and $f_{so}$ are the same function, but with different parameters, i.e.,
$f_{(so)}(r)=\{1+\exp[(r-r_{0(so)}A^{1/3})/a_{0(so)}]\}^{-1}$. For protons, the Coulomb potential
$V_{Coul}$ is taken to be that of a uniformly charged sphere with charge $(Z-1)e$ ($Z$ being the
total charge of the system) and  the radius $R_c$, which is added to
expression (\ref{eq_15}).

We will present numerical results for spherical heavy (stable) nuclei like
$^{132}$Sn, $^{208}$Pb and $^{298}$114 as well as for drip-line nuclei like
$^{154}$Sn and $^{180}$Pb. The potential parameters along with the maximal
number of nodes $n_{max}$ and partial waves $l_{max}$ included in the harmonic oscillator basis for converged values of the
sp energies are given in Table 1. In a realistic calculation, the potential parameters should be
selected in such a way that the experimental sp energies around the 
Fermi level are reproduced.

\begin{table}
\caption{Potential parameters used to calculate the bound states: The potential depth is
given in MeV, the radii and diffusenesses in fm, while the strength $\lambda$ of the spin-orbit
potential is adimensional.
See text for further details.}
\begin{ruledtabular}
\begin{tabular}{cccccccccc}
\multicolumn{10}{c}{Protons} \\
Nucleus& $V_0$ & $r_0$ & $a_0$ & $\lambda$ & $r_{so}$ & $a_{so}$ & $R_c$ & $n_{max}$
&$l_{max}$ \\
\hline
$^{132}$Sn$^{b}$ & 59.94 & 1.275 & 0.70  & 36 & 1.30 & 0.70 & 6.62 & 10 & 7 \\
$^{208}$Pb$^{a}$ & 59.21 & 1.286 & 0.762 & 18.29 & 0.88 & 0.60 & 6.69 & 10 & 7 \\
$^{298}$114$^{b}$ & 59.62 & 1.275 & 0.70 & 36 & 1.30 & 0.70 & 8.86 & 12 & 10 \\
$^{180}$Pb$^{b}$ & 53.39 & 1.275 & 0.70 & 36 & 1.30 & 0.70 & 7.34 & 10 & 7 \\
\multicolumn{10}{c}{Neutrons} \\
$^{132}$Sn$^{b}$ & 39.26 & 1.347 & 0.70 & 36 & 1.30 & 0.70 &  & 10 & 7 \\
$^{208}$Pb$^{c}$ & 44 & 1.27 & 0.67 & 32 & 1.27 & 0.67 &  & 10 & 7 \\
$^{298}$114$^{c}$ & 43 & 1.27 & 0.67 & 32 & 1.27 & 0.67 &  & 12 & 10 \\
$^{154}$Sn$^{b}$ & 34.64 & 1.347 & 0.70 & 36 & 1.30 & 0.70 &  & 10 & 7 \\
\end{tabular}
\\$^{a}$From Ref. \cite{Pruess}, $^{b}$From Ref. \cite{Dudek}, $^{c}$From Ref. \cite{Ross}
\end{ruledtabular}
\end{table}

\begin{figure}
\begin{center}
\includegraphics{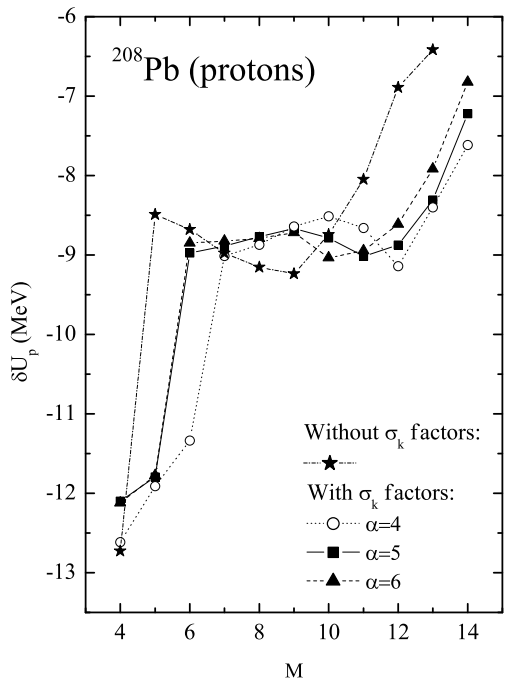}
\end{center}
\caption{Optimal proton shell correction $\delta U_{p}$ for $^{208}$Pb as a 
function of the degree $M$
of the smoothing function $\zeta_M$. The lines are to guide the eyes. 
See text for further details.}
\end{figure}

Fig. 1 shows the optimal proton shell corrections $\delta U_p$
for $^{208}$Pb as a function of the $M$ degree of the smoothing function $\zeta_M$.
The dependence of
$\delta U_p$ on the $\alpha$ parameter of the $\sigma_k$ factors
[$\alpha = 4$ (circle), $5$ (squares) and $6$ (triangles)] as well as
the effect of these factors on the plateau formation of $\delta U_p$ (e.g., comparing
squares with stars) can also be seen in Fig. 1.
It is observed that the best plateau is formed for $\alpha = 5$, and the
Lanczos factors are hugely important to determine a unique value of the shell correction. As in Ref. \cite{Vertse2}, it was found that the values of
$\Delta_{op}$ are strongly correlated with $M$, i.e., in the plateau of
$\delta U$ $\Delta_{op}$ increases with increasing $M$.

\begin{figure}
\begin{center}
\includegraphics{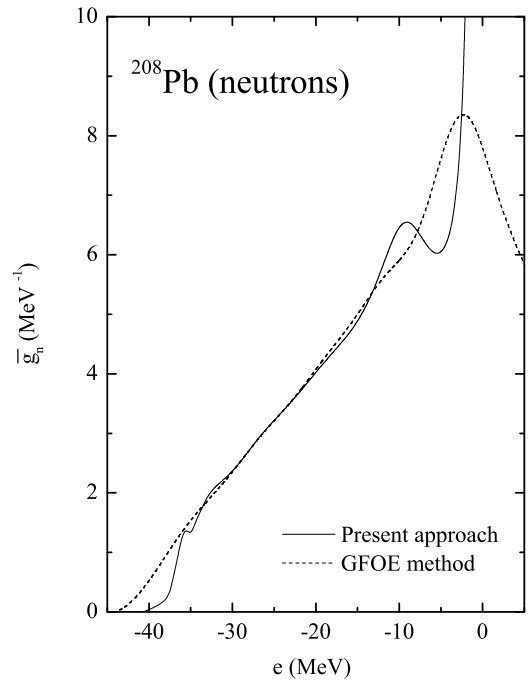}
\end{center}
\caption{The smoothed sp level density of the neutrons in $^{208}$Pb, 
resulting from
the present approach (solid curve), is compared to that obtained by us with 
the GFOE method (dashed curve). See text for further details.}
\end{figure}

In Fig 2, we compare the smoothed sp level density of the neutrons in $^{208}$Pb, which results from
expressions (\ref{eq_10c})-(\ref{eq_10d}) for $M = 10$ at $\Delta_{op}^{n} = 0.99$ (solid curve),
to that obtained by us with the GFOE method \cite{Vertse2} (dashed curve).
The number
of oscillator shells included in the basis for the GFOE-calculation is
$N_{osc}=36$, the degree of the smoothing polynomial is 10, and the
resulting optimal energy smoothing parameter
$\gamma_{op} = 1.57\hbar\omega_0$.
The smoothed sp
level densities are very similar to each other
in the middle of the energy region where their linearity is required by the generalized Strutinsky
smoothing condition. The remaining oscillations of the solid curve indicate that the effect of the
shell fluctuations has not yet been completely eliminated by the found value of $\Delta_{op}^{n}$.
In contrast to the dashed curve, the solid curve abruptly increases near the continuum threshold
and this behaviour is the same as that of the semiclassical level density obtained with the
Wigner-Kirkwood method (see Ref. \cite{Vertse1}). It is important to stress that unlike the
$E$-averaging procedures such as the GFOE method or the semiclassical method, in the approach
suggested in the present work, which is based on the $N$-averaging, it is not
the smoothed sp level density that determines the shell corrections, but
only its linear behaviour in the intermediate
energy region.

\begin{figure}
\begin{center}
\includegraphics{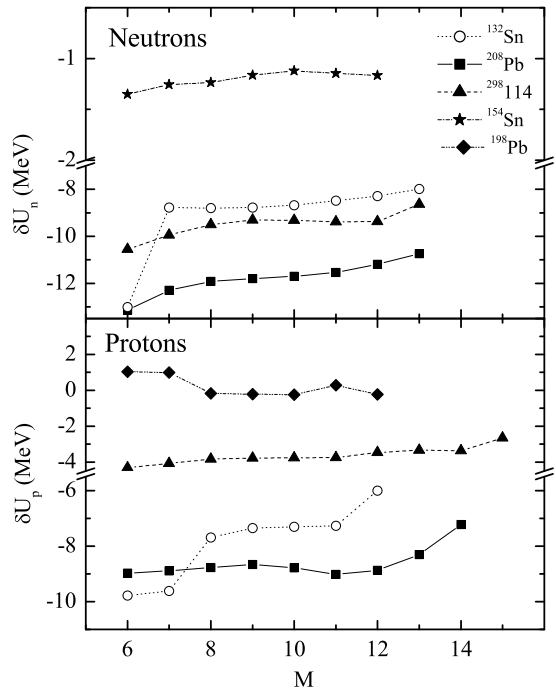}
\end{center}
\caption{Optimal neutron (upper part) $\delta U_{n}$ and proton (lower part) 
$\delta U_{p}$ shell corrections
for $^{132}$Sn (circles), $^{208}$Pb (squares), $^{298}$114 (triangles),
$^{154}$Sn (stars) and $^{180}$Pb (diamonds) as a function of the degree $M$
of the smoothing function $\zeta_M$. The lines are to guide the eyes. 
See text for further details.}
\end{figure}

\begin{table}
\caption{Shell corrections (in MeV) obtained with the present approach $\bar {\delta U}$ are compared to those
$\delta E$ calculated with the GFOE method \cite{Vertse2}. See text for further details.}
\begin{ruledtabular}
\begin{tabular}{ccccc}
&\multicolumn{2}{c}{Neutrons} & \multicolumn{2}{c}{Protons} \\
Nucleus& $\bar {\delta U_n}$ & $\delta E_n$ & $\bar {\delta U_p}$ & $\delta E_p$ \\
\hline
$^{132}$Sn & -8.70 & -9.18 & -7.40 & -7.10 \\
$^{208}$Pb & -11.74 & -11.00 & -8.85 & -8.12\\
$^{298}$114 & -9.37 & -8.40 & -3.78 & -3.90 \\
$^{154}$Sn & -1.20 & 9.30 & & \\
$^{180}$Pb & & & -0.22 & -8.0 \\
\end{tabular}
\end{ruledtabular}
\end{table}

Fig. 3 shows optimal neutron (upper part) and proton (lower part) shell corrections for
$^{132}$Sn (circles), $^{208}$Pb (squares), $^{298}$114 (triangles), $^{154}$Sn (stars)
and $^{180}$Pb (diamonds) as a function of the degree $M$ of the smoothing function $\zeta_M$.
A plateau is fairly well formed in all cases for $M$
around 10. The heavier the system (i.e., more sp levels are included), the better the plateau
can be seen. In table 2, the mean value of the optimal shell corrections belonging to the plateau
$\bar {\delta U}$  are compared to the shell corrections $\delta E$ obtained with the GFOE method.
The shell corrections of the present approach can be larger (e.g., for neutrons in $^{132}$Sn or for
protons in $^{298}$114) or smaller than those resulting from the GFOE
method.
The same features are observed comparing the shell corrections obtained
with the semiclassical Wigner-Kirkwood method and the GFOE method (see
Ref. \cite{Vertse1}). For well-bound nuclei, the maximal deviation found in table 2 is for
neutrons in
$^{298}$114 and is about 1 MeV. The difference cannot be explained by
the so-called
symmetry correction \cite{Ivanjuk2,Ivanjuk3} which is always positive and was found in
calculations to be about 2-3 MeV for the nuclei studied.
It seems that the generalized Strutinsky smoothing condition,
used here in combination with the N-averaging procedure,
removes large part of the difference between the shell corrections arising from the
$N$-averaging and the $E$-averaging prescriptions \cite{Ivanjuk2} for well-bound nuclei.
For the two drip-line nuclei studied,
namely $^{154}$Sn (neutrons) and $^{180}$Pb (protons), it was
found that the optimal shell corrections also shows a plateau with respect to $M$, but
in this case the plateau largely deviates from the shell correction provided by the GFOE method
(see table 2). Since reasons for such a discrepancy are
still unknown, we do not recommend the use of the present method to drip-line nuclei for the time
being. Further investigation is needed to clarify this point.

$\textit{Conclusions:}$ To conclude we would say that ground-state shell
corrections for finite depth potentials can be
calculated using bound states only, by means of the new method proposed in the present paper.
The N-averaging
procedure to obtain the smoothed total sp energy has been combined with both the generalized
Strutinsky smoothing
condition and the Lanczos $\sigma$ factors.
The new method provides unique
values of the shell corrections and
strongly reduces the well-known difference between the $N$-averaging and the $E$-averaging
prescriptions for well-bound (stable) nuclei, but this is not the case for drip-line nuclei.
The method can be useful to calculate potential energy surfaces for fusion and fission in the
macroscopic-microscopic approach with a two-center shell model based on two realistic Woods-Saxon
potentials without the need of including unbound sp levels.
Works in this direction are in progress.


$\textit{Aknowledgements:}$ The author thanks W. Scheid for fruitful discussions, and
the Alexander von Humboldt Foundation for financial support.

\end{document}